\documentclass[aps,prd,nofootinbib,showpacs,showkeys,preprint,floatfix]{revtex4}
\usepackage{amssymb}
\usepackage{amsmath}
\usepackage{amsfonts}
\usepackage{graphicx}
\usepackage{epsfig}
\usepackage{color}
\usepackage{xspace}
\usepackage{ulem}

\def\gsim{\raise0.3ex\hbox{$\;>$\kern-0.75em\raise-1.1ex\hbox{$\sim\;$}}}
\def\lsim{\raise0.3ex\hbox{$\;<$\kern-0.75em\raise-1.1ex\hbox{$\sim\;$}}}

\newcommand{\ba}[1]{\begin{eqnarray} \label{(#1)}}
\newcommand{\ea}{\end{eqnarray}}

\newcommand{\be}[1]{\begin{equation} \label{(#1)}} 
\newcommand{\ee}{\end{equation}}  
 
\def\lsim{\mbox{${}^< \hspace*{-7pt} _\sim$}} 
\def\gsim{\mbox{${}^> \hspace*{-7pt} _\sim$}}

\def\m{\mbox{$\mu^--e^-$}}

\newcommand{\AddrTU}{%
Institut f\"ur Theoretische Physik, Universit\"at T\"ubingen, \\
Kepler Center for Astro and Particle Physics, \\ 
Auf der Morgenstelle 14, D-72076 T\"ubingen, Germany}

\newcommand{\AddrUFSM}{
Universidad T\'ecnica Federico Santa Mar\'\i a, \\ 
Centro-Cient\'\i fico-Tecnol\'{o}gico de Valpara\'\i so, \\ 
Casilla 110-V, Valpara\'\i so,  Chile}

\def\gsim{\raise0.3ex\hbox{$\;>$\kern-0.75em\raise-1.1ex\hbox{$\sim\;$}}}
\def\lsim{\raise0.3ex\hbox{$\;<$\kern-0.75em\raise-1.1ex\hbox{$\sim\;$}}}

\begin{document}

\title{Limits on Lepton Flavor Violation from  $\mu^--e^-$-conversion}

\author{Marcela Gonz\' alez} 
\email{marcela.gonzalez@postgrado.usm.cl}\affiliation{\AddrUFSM}
\author{Juan Carlos Helo} 
\email{juan.heloherrera@gmail.com}\affiliation{\AddrUFSM}
\author{Sergey Kovalenko}\email{sergey.kovalenko@usm.cl}\affiliation{\AddrUFSM}
\author{Ivan Schmidt}\email{ivan.schmidt@usm.cl}\affiliation{\AddrUFSM}
\author{Thomas Gutsche} 
\email{thomas.gutsche@uni-tuebingen.de}\affiliation{\AddrTU} 
\author{Valery E. Lyubovitskij \footnote
{On leave of absence from Department of Physics, 
Tomsk State University, 634050 Tomsk, Russia}}
\email{valeri.lyubovitskij@uni-tuebingen.de}\affiliation{\AddrTU}

\keywords{lepton flavor violation, $\mu -e$ 
conversion, leptons, light and heavy mesons, 
contact interactions} 

\pacs{11.30.Fs, 12.60.-i, 14.60.Cd, 14.60.Ef, 23.40.Bw}

\begin{abstract}

We revisit the status of  Lepton Flavor Violating (LFV) 
$\mu e qq$ contact interactions from the view point of \m-conversion 
in nuclei. 
We consider their contribution to this LFV process  
via the two mechanisms on the hadronic level: 
direct nucleon and meson exchange ones. 
In the former case the quarks are embedded directly  into the nucleons while 
in the latter in mesons which then interact with nucleons in a nucleus.  
We revise and in some cases reevaluate the hadronic  parameters relevant 
for both mechanisms and calculate the contribution of the above mentioned 
contact interactions in coherent \m-conversion in various nuclei.  
Then we update our previous upper bounds and derive new ones for  
the scales of the $\mu e qq$ contact interactions from the experimental 
limits on the capture rates of \m-conversion. 
We compare these limits with the ones 
derived in the literature from other LFV processes and comment on the 
prospects of LHC searches related to the contact $\mu e qq$ interactions. 

\end{abstract}

\maketitle

\section{Introduction}
\label{Introduction}

Neutrino oscillations give the first and so far unique evidence for 
Lepton Flavor Violation~(LFV), forbidden in the Standard Model~(SM). 
Knowing that {\it lepton flavor} is a non-conserving quantity, 
it is natural to expect LFV effects also in the sector of charged leptons, 
although so far these effects have not been experimentally observed. 
Theoretically LFV in the neutrino sector, originating from the non-diagonal 
neutrino mass matrix,  is transmitted to  the charged lepton sector  
at the loop-level, in the form of penguin and box diagrams with virtual 
neutrinos.  However, these effects are Glashow-Iliopoulos-Maini (GIM)-like 
suppressed down to the level of $10^{-50}$, being far beyond the experimental 
reach. On the other hand, the charged lepton sector may receive other 
LFV contributions from physics beyond the SM, attributed to a certain 
high-energy scale $\Lambda_{LFV}$, which is not a priori necessarily  
very  high and which may provide observable LFV phenomena.

Thus, searching for lepton flavor violation in reactions with 
charged leptons offers a good opportunity for getting information on 
possible physics beyond the SM. 
Muon-to-electron conversion in nuclei 
\begin{eqnarray} 
\mu^- + (A,Z) \longrightarrow  e^- \,+\,(A,Z)^\ast 
\label{I.1} 
\end{eqnarray} 
is well known  to be one of the most sensitive probes of LFV and of underlying 
physics beyond the SM (for reviews, 
see~\cite{Marciano:2008zz,Kuno:1999jp,Czarnecki:2002ac,Kosmas:1993ch}).
Up to now there have been undertaken significant efforts aimed at searching 
for LFV via  this processes  in various nuclei with negative 
results~\cite{Marciano:2008zz}, thus setting upper limits on the \m-conversion rate
\begin{eqnarray}\label{Ti}  
&&R_{\mu e}^{A} = \frac{\Gamma(\mu^- + (A,Z) \rightarrow e^- + 
(A,Z))} {\Gamma(\mu^- + (A,Z)\rightarrow \nu_{\mu} + (A,Z-1))}\,.  
\end{eqnarray} 
The SINDRUM II experiment at PSI has set stringent upper bounds on 
\m-conversion rate $R_{\mu e} \leq 4.3\times 10^{-12},\  
7.0\times 10^{-13},\ 4.6\times 10^{-11}$ in  $^{48}$Ti \cite{Dohmen:1993mp},  
$^{197}$Au \cite{Bertl:2006up} and $^{208}$Pb \cite{Honecker:1996zf} 
as stopping targets respectively. 
Several new proposals for \m-experiments are aimed at a significant improvement 
of the SINDRUM II sensitivity. Among them  we mention the planned nearest 
future DeeMe experiment at J-PARC~\cite{Aoki:2010ve},  the next generation 
muon-to-electron conversion experiment by Mu2e Collaboration at 
Fermilab~\cite{Carey:2008zz,Kutschke:2011ux} and 
COMET at J-PARC~\cite{Cui:2009zz} with planned sensitivities 
around $10^{-14}$, $7\times 10^{-17}$ and $10^{-16}$ respectively, 
as well as the more distant future proposal PRISM/PRIME~\cite{Witte:2012zza} 
at J-PARC, with estimated sensitivity $10^{-18}$. 

As is known from previous studies (see, for instance, 
Refs.~\cite{Marciano:2008zz,Czarnecki:2002ac,Faessler:2004ea} 
and references therein) and as will be also discussed later in the 
present paper, these experimental bounds allow setting stringent bounds 
on the mechanisms of \m conversion~\cite{Faessler:2004ea}, on LFV 
decays of vector mesons~\cite{Gutsche:2009vp,Gutsche:2011bi} and, in general, on the 
underlying theories of LFV \cite{Czarnecki:2002ac}. 

The theoretical studies of \m conversion, presented in the literature, 
cover various aspects of this LFV process: the adequate treatment 
of structure effects~\cite{Kosmas:1993ch,Kosmas:2001mv,Faessler:2000pn}  
of the nucleus participating in the reaction and the underlying 
mechanisms of LFV at the quark level within different scenarios of 
physics beyond the SM (see~\cite{Czarnecki:2002ac} and references therein). 

As is known there are two categories of  \m conversion mechanisms: photonic and 
non-photonic. In the photonic case  
photon connects the LFV leptonic and 
the electromagnetic nuclear vertices.  
The non-photonic mechanisms are induced by the four-fermion 
lepton-quark LFV contact interactions. These mechanisms
significantly differ from each other, receiving different 
contributions from new physics and requiring different description of 
the nucleon and the nuclear structure. 

In the present paper we analyze  non-photonic mechanism. We revisit some of the 
results of Refs.~\cite{Faessler:2004ea,Faessler:2004jt,Faessler:2005hx} using 
improved values of hadronic parameters.  Then we significantly extend our previous analysis 
made in these papers  including all the possible contact terms contributing to \m conversion. 
To this end we evaluate nucleon form factors for the heavy quark currents and
take into account contributions of heavy vector mesons. 
Finally we update our previous  bounds on the $\mu e q q$ 4-fermion contact interactions and derive new ones
from the experimental data on \m  conversion rates in various nuclei. 

The paper is organized as follows. In Sec.~II we specify all the above 
mentioned LFV contact interactions contributing to coherent \m-conversion 
and briefly describe their hadronization 
%leading to two mechanisms: 
within direct nucleon and meson exchange mechanisms. 
%In the former case the quarks are embedded directly into the nucleons 
%while in the latter one in mesons which then interact with nucleons 
%in a nucleus. 
In Sec.~III  we consider the existing and future \m-conversion data and
%we show and discuss the main formula for the \m-conversion 
%capture rate as well as all the necessary atomic and nuclear structure 
%parameters involved in it. 
%Then 
extract   limits on the generic LFV parameters and on the equivalent 
mass scales $\Lambda_{\mu e}^{(q)}$ of the $\mu e qq$ contact interactions.  
Then we compare our limits with the existing ones in the literature and comment on 
a possible experimental reach of the LHC experiments in terms of  the mass scales of these interactions.  

\section{Model independent framework}

The effective Lagrangian ${\cal L}_{eff}^{lq}$ describing 
the coherent $\mu^{-}-e^{-}$-conversion at the quark level can be written in the form~\cite{Kosmas:2001mv,Faessler:2004jt}:
\begin{eqnarray} 
\label{eff-q}
{\cal L}_{eff}^{lq}\  &=& \  \frac{1}{\Lambda_{LFV}^2}  
\biggl[ \, \left(\eta_{VV}^{(q)} j_{\mu}^V\ + \eta_{AV}^{(q)} 
j_{\mu}^A \right)J_{q}^{V\mu}  
+ \left(\eta_{SS}^{(q)} j^S\ + \eta_{PS}^{(q)} j^P\ \right)J_{q}^{S} \biggr], 
\end{eqnarray} 
where 
the lepton and the quark currents are defined as   
\begin{eqnarray}\label{l-currents}
&&j_{\mu}^V 
= \bar e \gamma_{\mu} \mu\,, \,\,\,  
j_{\mu}^A = \bar e \gamma_{\mu} \gamma_{5} \mu\,, \,\,\,  
j^S = \bar e \ \mu\,, \\ \nonumber
&& j^P = \bar e \gamma_{5} \mu\,, \,\,\,  
J_{q}^{V\mu} = \bar q \gamma^{\mu} q\,, \,\,\, 
J_{q}^{S} = \bar q \ q \,.
\end{eqnarray}   
In Eq.~(\ref{eff-q}) the summation is understood 
over all the quark flavors $q= \{u,d,s,b,c,t\}$.
The dimension-1 mass parameter $\Lambda_{LFV}$  
is  a high energy scale of LFV connected 
to new physics. The dimensionless LFV parameters $\eta^{q}$ in 
Eq.~(\ref{eff-q}) refere to a specific LFV model. 
We start with studying  $\mu^{-}-e^{-}$-conversion in a model 
independent way based on the Lagrangian (\ref{eff-q}) and will extract 
upper limits on the parameters $\eta^{q}$ from the experimental bounds 
on this LFV process. We consider the dominant coherent mode of 
$\mu^{-}-e^{-}$-conversion, therefore, in Eq. (\ref{eff-q}) 
we disregarded the terms with the axial-vector and pseudoscalar quark 
currents which are irrelevant in this case  \cite{Kosmas:1993ch,Kosmas:2000nj}.

\begin{figure}[tbh]
\vspace{-30mm}
\includegraphics[width=12cm,height=14cm,angle=0]{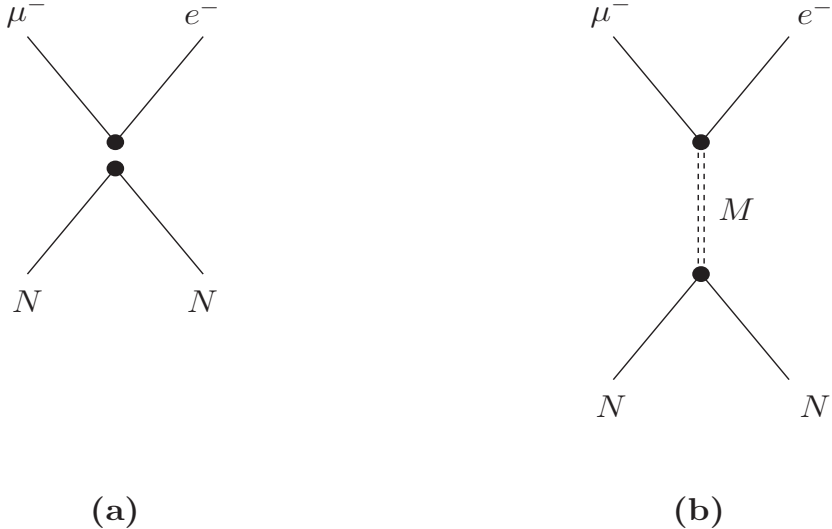}
\vspace{-35mm}
\caption{Diagrams contributing to the nuclear $\mu^--e^-$ conversion: 
direct nucleon (a) and meson-exchange mechanism (b)}
\label{Fig1}
\end{figure}

To this end one needs to translate the LFV lepton-quark 
Lagrangian~(\ref{eff-q}) to the corresponding 
LFV lepton-nucleon Lagrangian. This implies   
a certain hadronization prescription. Due to the absence of a well-defined theory 
of hadronization we rely on some reasonable assumptions and models. 
Following Refs.~\cite{Faessler:2005hx,Faessler:2004jt}
we consider two mechanisms of nuclear $\mu^- - e^-$ conversion: 
direct nucleon mechanism (DNM) and  vector-meson exchange mechanism  (MEM), 
which are shown in~Fig.~\ref{Fig1}.

%\begin{figure}
%\begin{center}
%\vspace*{1cm}
%\epsfig{figure=fig1.eps,scale=1}
%\end{center}
%\caption{Diagrams contributing to the nuclear $\mu^--e^-$ conversion: 
%direct nucleon mechanism (a) and 
%meson-exchange mechanism (b)}
%\label{Fig1}
%\end{figure}

In the case of the DNM  the quark currents are directly embedded  into the 
nucleon currents (Fig.1a).
The MEM consists of two stages (Fig.1b). First, the   
quark currents are embedded into the interpolating meson fields which
then interact with the nucleon currents. It is well known that only 
vector~\cite{Faessler:2004ea,Faessler:2004jt} and 
scalar~\cite{Faessler:2005hx} mesons contribute to coherent 
$\mu-e$-conversion while axial and pseudoscalar mesons contribute to
a subdominant incoherent channel of this process.  
%
%{SK_22.03.2013} Modif
%In general one expects all the mechanisms to contribute 
%to $\mu-e$-conversion. However, at present the relative 
%strengths for the different mechanisms are unknown. 
%In view of this problem we assume, for the first approximation, 
%that only one mechanism is operative and estimate its contribution 
%to the process in question. 
%
At present the relative 
strengths for the DNM and the MEM mechanisms cannot be reliably determined. 
Therefore, in our analysis  we assume
%, for the first approximation, 
that only one of the  mechanism is operative at a time and estimate its contribution 
to $\mu^- - e^-$ conversion. 
%{SK_22.03.2013} END

\subsection{Direct Nucleon Mechanism}

In the case of the DNM,  schematically represented by the diagram in Fig.1(a), 
the quark fields from Eq. (\ref{eff-q}) are embedded in 
the effective nucleon fields $N$. 
Then the effective Lagrangian  of $\mu^- - e^-$ conversion can be written 
in a general Lorentz covariant form with the isospin structure of the 
$\mu-e$-transition operator~\cite{Kosmas:2001mv}:  
%{SK_22.03.2013} Typo corrected.
\begin{eqnarray} 
\hspace*{-.5cm} 
{\cal L}_{eff}^{lN} =   \frac{1}{\Lambda_{LFV}^2} 
\left[ j_{\mu}^h \left(\alpha_{hV}^{(0)} J^{V\mu \, (0)} + 
\alpha_{hV}^{(3)} J^{V\mu \, (3)}\right) + j^r \left(\alpha_{r S}^{(0)} 
J^{S \, (0)} + \alpha_{r S}^{(3)} J^{S \, (3)}\right)\right]\,, 
\label{eff-N} 
\end{eqnarray} 
%{SK_22.03.2013} END
where the summation runs over the double indices $h = V,A$ and 
$r = S,P$. The isosinglet $J^{(0)}$ and isotriplet $J^{(3)}$ 
nucleon currents are defined as 
\begin{eqnarray}\label{currents-2}
J^{V\mu \, (k)} \, = \, \bar N \, \gamma^\mu \, \tau^k \,  N, \ \ \  
J^{S \, (k)}  \, = \, \bar N \, \tau^k \, N\,,
\end{eqnarray} 
where $N$ is the nucleon isospin doublet, $ k = 0,3 $ and 
$\tau_0\equiv\hat I$ is the $2\times 2$ unit matrix.  

This Lagrangian is supposed to originate from the quark level Lagrangian in  
Eq.~(\ref{eff-q}), and therefore must correspond to the same  
order $1/\Lambda_{LFV}^{2}$ in inverse powers of the LFV scale.

Next we relate the coefficients $\alpha$ in Eq.~(\ref{eff-N}) to the 
"fundamental"  LFV parameters $\eta$ of the quark level 
Lagrangian (\ref{eff-q}). Towards this end we apply the on-mass-shell 
matching condition~\cite{Faessler:1996ph}
\begin{equation}
\langle e^-N|{\cal L}_{eff}^{q}|\mu^-N\rangle \approx
\langle e^-N|{\cal L}_{eff}^{N}|\mu^-N\rangle ,  
\label{match}
\end{equation}
in terms of the matrix elements of the Lagrangians (\ref{eff-q}) and 
(\ref{eff-N}) between the initial and final states of \m conversion at 
the nucleon level.

In order to solve this equation in Ref.~\cite{Kosmas:2001mv}, various 
relations for the matrix elements of quark operators between nucleon 
states were used 
\begin{eqnarray}\label{mat-el1}
\langle N|\bar{q}\ \Gamma_{K}\ q|N\rangle = G_{K}^{(q,N)}
\bar{N}\ \Gamma_{K}\ N,
\end{eqnarray}
with $q=\{u,d,s\}$,  $N=\{p,n\}$. 
In the following we 
neglect the $q^2$-dependence of the nucleon form factors 
$G_{K}^{(q,N)}$, because the maximal  momentum transfer 
in $\mu -e$ conversion is  significantly  smaller than the 
typical scale of nucleon structure $\Lambda \sim 1$ GeV. 
Following Ref. ~\cite{Kosmas:2001mv} we find  relations between the 
LFV parameters of the Lagrangians (\ref{eff-q}) and (\ref{eff-N}):
\begin{eqnarray} 
\label{DNM-alpha-V}
\hspace*{-.65cm}
\mbox{DNM:}&&\alpha_{hV}^{(3)} = \frac{(G_{V}^{u} - G_{V}^{d})}{2} 
\eta_{hV}^{(3)}\,, 
\,\,\,  
\alpha_{hV}^{(0)}  = \frac{(G_{V}^{u} + G_{V}^{d})}{2} \eta_{hV}^{(0)},\\
&&\alpha_{r S}^{(3)} = \frac{(G_{S}^{u} - G_{S}^{d})}{2} \eta_{r S}^{(3)}\,, 
\,\,\,  
\alpha_{r S}^{(0)}  = \frac{(G_{S}^{u} + G_{S}^{d})}{2} 
\eta_{r S}^{(0)} + \eta^{(s)}_{r S} G^{s}_{S} \nonumber\\
&& \hspace{51mm} +\  \eta^{(c)}_{r S} G^{c}_{S}  
+  \eta^{(b)}_{r S} G^{b}_{S} +  \eta^{(t)}_{r S} G^{t}_{S}\,, 
\label{DNM-alpha-S}
\end{eqnarray} 
%where $h=V,A$ and $f=S,P$.  
where $h = V, A$, $r=S,P$ and $\eta^{(0,3)}=\eta^{(u)}\pm\eta^{(d)}$.

Here the nucleon form factors have the following 
values~\cite{Kosmas:2001mv,Faessler:2005hx,Kosmas:2001sx}:
\begin{eqnarray}\label{FFN}
G^{u}_{V} =2; \ \  
G^{d}_{V} =1;\ \  
G^{u}_{S} &=& 3.74 [5.1]; \ \ 
G^{d}_{S} =2.69 [4.3]; \ \ G^{s}_{S} = 0.64 [2.5]; 
\\
\label{FFN-h}
G^{c}_{S} &=& 0.06; \ \ \ \ \ \ \  \  
G^{b}_{S} = 0.02;\ \  \ \ \ \ \  
G^{t}_{S} = 5\times 10^{-4}\,
\end{eqnarray}
The values of the vector form factors $G_{V}^{q}$ are exact and are equal to 
the total number of the corresponding specie  $q$ of quark
in the proton.  For this reason $G_{V}^{s,c,b,t}= 0$.  
For the scalar form factors $G^{u,d,s}_{S}$ we use the conservative 
values derived in Ref.~\cite{Faessler:2005hx}. 
They are deduced from the values of the meson-nucleon sigma terms of 
Refs.~\cite{Gasser:1990ce,Gasser:2000wv, Sainio:2001bq} which are extracted 
from the data on the basis of dispersion analysis of $\pi N$ scattering 
data taking into account chiral symmetry constraints. 
In the square brackets we also show the significantly larger values of 
the scalar form factors derived in Ref.~\cite{Kosmas:2001mv} within the 
QCD picture of baryon masses as based on~\cite{Shifman:1978zn,Cheng:1988cz}. 
The latter approach also allows for an estimate of the heavy quark scalar 
form factors $G_{S}^{c,b,t}$ of the nucleon.  The value of $G_{S}^{b}$ was 
calculated in Ref.~\cite{Kosmas:2001sx}. Here we calculated in the same 
approach values of the remaining scalar form factors  $G_{S}^{c}$ and 
$G_{S}^{t}$. The resulting values  of  $G_{S}^{c,b,t}$ are shown in 
Eq.~(\ref{FFN-h}). 
For a discussion of theoretical  
uncertainties and the possible error bars see, for instance, 
Ref.~\cite{Corsetti:2000yq}.

\subsection{Meson Exchange Mechanism}

This mechanism is described by the diagram in Fig.1(b).  
As we already mentioned, the mesons that can contribute to this mechanism 
are the unflavored vector and scalar  ones.  
%The lightest  give the dominant contribution to \m-conversion, and 
%these 
The lightest vector mesons are the  isotriplet $\rho(770)$ and 
the two isosinglet  $\omega(782), \ \phi(1020)$ mesons.
In our analysis we adopt the ideal singlet-octet mixing, corresponding to 
the following quark content of the $\omega$ and $\phi$ 
mesons~\cite{Particle-Data-Group:2012uq}:
%\begin{eqnarray}\label{quark-cont}
$\omega = (u \bar u + d \bar d)/\sqrt{2}, \ \ \ \phi = - s \bar s$.
%\end{eqnarray} 
 
Contributions of the heavy vector mesons $J/\Psi$ and $\Upsilon$ 
are significantly suppressed in comparison with the above specified light 
vector mesons but these mesons probe the heavy quark vector currents of 
the nucleon inaccessible in the DNM and therefore are worth to be taken 
into account.

The properties of  scalar mesons are not yet well experimentally 
established~\cite{Particle-Data-Group:2012uq}. However their phenomenological 
role in the \m conversion could be important, because they contribute as well 
as the vector mesons to the experimentally most interesting coherent mode 
of this rare process.  
The isosinglet $f_0(500)$ and the isotriplet $a_0(980)$ states 
are the lightest unflavored scalar mesons. 
The $f_0(500)$ meson has been considered in the context of the nonlinear 
realization of chiral symmetry as a wide resonance 
in the $\pi\pi$ system (see e.g. in 
Refs.~\cite{Colangelo:2001df,Oller:1997ng,Oset:2000gn}). 
There should also be mentioned a model-independent study of scalar mesons using 
uniformizing-variable method based on analyticity and unitarity of 
the $S$-matrix~\cite{Surovtsev:2011yg}. 
In our analysis we neglect a possible small strangeness content of 
the isosinglet meson and take it in the form: $f_0(500) = (\bar u u + \bar d d)/\sqrt{2}$.  

The upper vertex of diagram Fig.1(b) is described by the 
LFV effective lepton-meson Lagrangian~\cite{Faessler:2004jt}:  
\begin{eqnarray}\label{eff-LV}
{\cal L}^{lM} &=& \frac{\Lambda_H^2}{\Lambda_{LFV}^2}  
\biggl[ \, (\xi_V^{M_{V}} j_{\mu}^V\ +
                    \xi_A^{M_{V}} j_{\mu}^A\ ) M_{V}^{\mu} 
+    ( \xi_S^{M_{S}} j^S \, + \, \xi_P^{M_{S}}j^P ) \, M_{S} 
\nonumber \\
&+&\frac{1}{\Lambda_H^2}
\biggl\{{\rm Derivative \  terms} \biggr\}\biggr]\,, 
\end{eqnarray}  
with $M_{V}= \rho, \omega, \phi, J/\Psi, \Upsilon$ and 
$M_{S} = f_{0}, a_{0}$ mesons. 
The unknown dimensionless coefficients $\xi$ are  to be determined 
from the had\-ro\-ni\-za\-ti\-on prescription. Since we suppose that 
this Lagrangian originates from the quark-lepton Lagrangian (\ref{eff-q}),  
all its terms are suppressed by a factor $\Lambda_{LFV}^{-2}$ with 
respect to the large LFV mass scale $\Lambda_{LFV}$. 
Another mass scale in the 
problem is the hadronic scale $\Lambda_H \sim 1$ GeV. It is introduced in 
the Lagrangian of Eq.~(\ref{eff-LV}) in order to adjust 
physical dimensions of its terms. 
Typical momenta involved in $\mu^- - e^-$ conversion are $q \sim m_{\mu}$ 
where $m_{\mu}$ is the muon mass. 
Thus, from naive dimensional counting 
one expects that the contribution of the derivative terms to 
$\mu^- - e^-$ conversion is suppressed by a factor 
$(m_{\mu}/\Lambda_H)^2\sim 10^{-2}$ in comparison with the non-derivative terms. Therefore, 
%at this step in 
we retain in Eq.~(\ref{eff-LV})  only the dominant non-derivative terms. 
For a more detailed discussion of the role of the derivative terms see
Ref. ~\cite{Faessler:2004jt}.

We relate the parameters of Lagrangians (\ref{eff-LV}) 
and (\ref{eff-q}) with the help of   
%by analogy as in the previous section, 
%an approximate method based 
the on-mass-shell matching condition proposed in Refs.~\cite{Faessler:2004ea,Faessler:2004jt,Faessler:2005hx}:
\begin{equation}\label{match1} 
\langle \mu^+ \, e^-|{\cal L}_{eff}^{lq}|M\rangle \approx 
\langle \mu^+ \, e^-|{\cal L}_{eff}^{lM}|M \rangle ,  
\end{equation} 
with $|M= \rho, \omega, \phi, J/\Psi, \Upsilon, a_{0}, f_{0} \rangle$ 
corresponding to meson states on their mass-shells. 
This equation can be solved using the 
well-known quark current matrix elements for vector and scalar mesons 
\begin{eqnarray}\label{mat-el2} 
&&\langle 0|\bar u \, \gamma_\mu \, u|\rho^0(p,\epsilon)\rangle \, = \, 
\, - \, \langle 0|\bar d \, \gamma_\mu \, d|\rho^0(p,\epsilon)\rangle 
= \, m_{\rho}^2 \, f_{\rho} \, \epsilon_\mu(p)\,,\\ 
&&\langle 0|\bar{u}\ \gamma_{\mu}\ u|\omega(p,\epsilon)\rangle\hspace{0.7mm} = 
\hspace{5mm} \langle 0|\bar{d}\ \gamma_{\mu}\ d|\omega(p,\epsilon)\rangle 
\hspace{4.8mm} = \, 3 \, m_{\omega}^2 \, f_{\omega} \, \epsilon_\mu(p)\,,\\ 
&&\langle 0|\bar{s}\ \gamma_{\mu}\ s|\phi(p,\epsilon)\rangle 
\hspace{2mm} = \, - \, 3 \, m_\phi^2 \, f_{\phi} \, \epsilon_\mu(p) \,,\\ 
&&\langle 0|\bar{c}\ \gamma_{\mu}\ c|J/\Psi(p,\epsilon)\rangle 
\hspace{2mm} = \,  m_{J/\Psi}^2 \, f_{J/\Psi} \, \epsilon_\mu(p) \,, \\ 
&&\langle 0|\bar{b}\ \gamma_{\mu}\ b|\Upsilon(p,\epsilon)\rangle 
\hspace{2mm} = \,  m_{\Upsilon}^2 \, f_{\Upsilon} \, \epsilon_\mu(p) \,,\\
&&\langle 0|\bar u \, u|f_0(p)\rangle \, = \, 
 \langle 0|\bar d \, d|f_0(p)\rangle 
\, = \, m_{f_0}^2 \, f_{f_0}\,, \label{f0-dec}\\
&&     \langle 0|\bar u \, u|a_0^0(p)\rangle = 
\,- \, \langle 0|\bar d \, d|a_0(p)\rangle 
\, = \, m_{a_0}^2 \, f_{a_0} \,. 
\label{a0-dec} 
\end{eqnarray} 
Here $p$, $m_M$  and $f_{M}$ are the 4-momentum, mass and dimensionless 
decay constant of the meson $M$, respectively, $\epsilon_{\mu}$ is the vector 
meson polarization state vector. 

The current central values of the meson decay constants $f_{V}$ 
and masses $m_V$ are~\cite{Particle-Data-Group:2012uq}:
\begin{eqnarray}\label{constants}  
&&f_{\rho}   = 0.2, \ \ 
 f_{\omega} = 0.059, \ \ 
 f_{\phi}   = 0.074, \ \    f_{J/\Psi} = 0.134,  \ \  f_{\Upsilon} = 0.08, 
\\   
&&m_{\rho}   = 771.1 \,\,  \mbox{MeV}, \,\,\,
 m_{\omega} = 782.6 \,\, \mbox{MeV}, \,\,\,  
 m_{\phi}   = 1019.5 \,\, \mbox{MeV},\\
&& m_{J/\Psi} = 3097\,  \mbox{MeV}, \ \ m_{\Upsilon} = 9460\, \mbox{MeV}, 
\ \  m_{f_0} = 500 \,\,  \mbox{MeV}, \ \ m_{a_0} = 984.7 \,\, \mbox{MeV} \,.   
\end{eqnarray} 

The decay constants $f_{f_{0}}$  and $f_{a_{0}}$  
in Eqs.~(\ref{f0-dec}), (\ref{a0-dec}) are not yet known experimentally. 
%In such case one  can try to estimate them within certain well grounded 
%theoretical approaches. 
In Ref.~\cite{Faessler:2005hx} we evaluated them 
on the basis of the linear $\sigma$-model in the case of $f_0$ 
meson~\cite{Delbourgo:1993dk,Delbourgo:1998ji} and with the help of 
the QCD sum rules for $a_0$ meson ~\cite{Maltman:1999jn}.  
The result is \cite{Faessler:2005hx}: 
\begin{eqnarray}
\label{FScalar}
f_{f_0} =  0.28,\ \ \ \ \  f_{a_0} \, = \, 0.19 \,. 
\end{eqnarray} 

Following Ref.~\cite{Faessler:2004jt,Faessler:2005hx} we find the solution of Eq. (\ref{match1})
in the form
%The result is  
\begin{eqnarray}
&&\xi_h^{\rho} \, = \, \left(\frac{m_{\rho}}{\Lambda_H}\right)^2 
f_{\rho} \,\eta_{h V}^{(3)}\,, \,\, \ \ 
\xi_h^{\omega} \, = \, 3 \left(\frac{m_{\omega}}{\Lambda_H}\right)^2  f_{\omega} \, \eta_{h V}^{(0)}, \  \ \ \xi_h^{\phi} \, = \, 
- 3 \left(\frac{m_{\phi}}{\Lambda_H}\right)^2  
 f_{\phi} \, \eta_{h V}^{(s)}, \\ 
&&  \xi_h^{J/\Psi} \, = \, \left(\frac{m_{J/\Psi}}{\Lambda_H}\right)^2 
f_{J/\Psi} \,\eta_{h V}^{(c)}\,, \hspace*{39mm}  
\xi_h^{\Upsilon} \, =   \, \left(\frac{m_{\Upsilon}}{\Lambda_H}\right)^2  
\, f_{\Upsilon} \, \eta_{h V}^{(b)},\\ 
&& \xi_r^{a_0} \, = \, \left(\frac{m_{a_0}}{\Lambda_H}\right)^2 
f_{a_0} \,\eta_{r S}^{(3)}\,, \hspace*{47mm}  
\xi_r^{f_0} \, =   \, \left(\frac{m_{f_0}}{\Lambda_H}\right)^2  
\, f_{f_0} \, \eta_{r S}^{(0)}\,, 
\end{eqnarray} 
where as before  $h = V, A$, $r=S,P$ and $\eta^{(0,3)}=\eta^{(u)}\pm\eta^{(d)}$.

The lower vertex of the diagram in Fig.1(b) corresponds to the conventional 
strong isospin invariant effective 
Lagrangian~\cite{Weinberg:1968de,Mergell:1995bf,Kubis:2000zd}: 
\begin{eqnarray}\label{MN} 
&&{\cal L}^{MN}\, =  
\bar{N}\left( \frac{1}{2} g_{_{\rho NN}} \, 
\gamma^{\mu} \rho^{k}_{\mu} \tau^{k} \,  
+  \frac{1}{2} g_{_{M^{(0)}_{V} NN}} \, 
\gamma^{\mu} M^{(0)}_{V\, \mu}  + 
g_{_{a_0 NN}} \, a^{k}_0 \, \tau^{k} \,  + \, g_{_{f_0 NN}} \, f_0\right) N\,. 
\end{eqnarray} 
Here $M^{(0)}_{V} = \omega, \phi, J/\Psi, \Upsilon$ are the isosinglet vector mesons.
In this Lagrangian we again neglected the derivative terms, irrelevant for 
coherent $\mu^- - e^-$ conversion. For the light vector meson-nucleon 
couplings we use numerical values taken from an updated dispersive 
analysis~\cite{Mergell:1995bf,Meissner:1997qt,Faessler:2004jt} 
\begin{eqnarray}
\label{VN-couplings}
g_{_{\rho NN}}= 4.0\,, \,\, 
g_{_{\omega NN}} = 41.8, \ \  
-18.3 \leq g_{_{\phi NN}} \leq -0.24.
\end{eqnarray} 
In Ref.~\cite{Gutsche:2011bi} the couplings $g_{_{J/\Psi NN}}$ and  
$g_{_{\Upsilon NN}}$  were extracted  from the 
QCD analysis~\cite{Brodsky:1981qf,Chernyak:1983ej} of the existing 
data~\cite{Particle-Data-Group:2012uq} on the decay rates 
\mbox{$\Gamma(J/\Psi \rightarrow \bar{p}p)$} and 
$\Gamma(\Upsilon \rightarrow\bar{p}p)$. 
Their values are
\begin{eqnarray}\label{JPSI}
g_{_{J/\Psi}NN} = 1.6 \times 10^{-3}\,, \ \ \ 
g_{_{\Upsilon NN}} = 5.6\times 10^{-6}\,.
\end{eqnarray}

%{SK_23.03.2013} 
In the literature there are also estimations of the scalar meson-nucleon couplings. In our analysis we use:
% the following values:
%
\begin{eqnarray}\label{sNN} 
g_{_{f_0 NN}} \simeq  8.5[5.0]\,, \ \ \ \ 
g_{_{a_0 NN}} \simeq g_{_{f_0 NN}} \,. 
\end{eqnarray}
The first number is an empirical value of the scalar meson coupling $g_{_{f_0 NN}}$ to provide the needed 
intermediate range nucleon-nucleon attraction according to Ref.~\cite{Machleidt:1987hj}. 
%It was found in Ref.~\cite{Machleidt:1987hj}.
%
The last approximate relation was obtained in the chiral unitary approach 
in Ref.~\cite{Oset:2000gn}. The value of the scalar  $f_{0}$ meson coupling
calculated in this approach is given in square brackets.  For our analysis 
we consider the empirical values of Ref.~\cite{Machleidt:1987hj} as more 
reliable, but we will also show our results for the smaller values of 
Ref.~\cite{Oset:2000gn}. 

The meson-exchange contribution to \m-conversion corresponding to the 
diagram in Fig.1(b) is of second order in the Lagrangian 
${\cal L}^{lM} + {\cal L}^{MN}$.  
%We estimate this contribution only for the process 
%of the dominant coherent $\mu^- - e^-$ conversion. 
%As we already stressed, in this case 
Considering coherent \mbox{$\mu^- - e^-$} conversion
we ignore, as justified before, all  
derivative terms involving nucleon and lepton fields. 
Furthermore we neglect the kinetic 
energy of the final nucleus, the muon binding energy and the electron 
mass. In this approximation the squared momentum, transferred to the nucleus, has a constant 
value $q^2 \approx - m_{\mu}^2$ and the meson propagators contract to \mbox{$\delta$-functions}. 
Thus the meson exchange contributions in Fig. 1(b)
%In this case vector meson exchanges 
result in effective lepton-nucleon  \mbox{4-fermion} operators of the same type as in Eq.~(\ref{eff-N}). 
%Therefore, the meson exchange diagram in Fig.1(b) 
%results in a MEM  contribution to 
For the corresponding $\alpha$ 
parameters we find:
\begin{eqnarray} 
\label{MEM-alpha-V}
\mbox{MEM}:\ \ \ \alpha_{h V}^{(3)} &=& - \beta_{\rho}\eta_{h V}^{(3)},
\ \ \  
\alpha_{hV}^{(0)} = - \beta_{\omega} \eta_{h V}^{(0)}
- \beta_{\phi}\eta_{h V}^{(s)} - \beta_{J/\Psi}\eta_{h V}^{(c)} 
- \beta_{\Upsilon}\eta_{h V}^{(b)}, \\
\label{MEM-alpha-S}
\alpha_{r S}^{(3)} &=& \beta_{a_0}\eta_{r S}^{(3)}\, ,
\hspace*{6mm}  
\alpha_{r S}^{(0)} = \beta_{f_0}\eta_{r S}^{(0)}
\end{eqnarray}
with $h=V,A$, $r=S,P$ and the coefficients 
\begin{eqnarray} \label{beta}
\beta{_{\rho}} &=& \frac{1}{2} 
\frac{g_{_{\rho NN}} \, f_{\rho} \, m_{\rho}^2}{m_{\rho}^2 
+ m_{\mu}^2},\  \  \ \ \  \ \ \  \  \ \ \beta{_{\omega}} = \frac{3}{2} 
\frac{g_{_{\omega NN}} \, f_{\omega} \, m_{\omega}^2 }
{m_{\omega}^2 + m_{\mu}^2},\  \ \ \  \  \ \  \ \  \ \ \beta{_\phi} = -\frac{3}{2} 
\frac{g_{_{\phi NN}} \, f_{\phi} \, 
m_{\phi}^2 }{m_{\phi}^2 + m_{\mu}^2},\\
\beta{_{J/\Psi}} &=& \frac{g_{_{J/\Psi NN}} \, f_{J/\Psi} \, 
m_{J/\Psi}^2}{m_{J/\Psi}^2 
+ m_{\mu}^2}\,, \hspace*{6mm} 
\beta{_{\Upsilon}} = \frac{g_{_{\Upsilon NN}} \, 
f_{\Upsilon} \, m_{\Upsilon}^2 }
{m_{\Upsilon}^2 + m_{\mu}^2}, \\
\beta{_{a_0}} &=& \frac{g_{_{a_0 NN}} \, f_{a_0} \, 
m_{a_0}^2}{m_{a_0}^2 
+ m_{\mu}^2}\,, \hspace*{12.5mm} 
\beta{_{f_0}} = \frac{g_{_{f_0 NN}} \, f_{f_0} \, m_{f_0}^2 }
{m_{f_0}^2 + m_{\mu}^2}\,.
\end{eqnarray}
Substituting the values of the meson parameters 
from Eqs.~(\ref{constants})-(\ref{FScalar}) and (\ref{VN-couplings})-(\ref{sNN})
we obtain
% for these coefficients 
\begin{eqnarray} \label{beta-num} 
\beta{_{\rho}} &=& 0.39\,, \,\,\, 
\ \ \ \ \ \ \  \beta{_{\omega}} = 3.63, \ \ \  \ \ \ \ 
0.03 \leq \beta{_{\phi}} \leq 2.0, \\
\beta_{_{J/\Psi}} &=& 2\times 10^{-4}, \ \ \ 
\beta_{_{\Upsilon}} = 5 \times 10^{-7}, \\
\beta{_{a_0}} &=& 1.58[0.93] \,, \hspace*{1.5mm} 
\beta{_{f_0}} = 2.24[1.32].
\end{eqnarray} 
The coefficients $\beta{_{a_0}}, \beta{_{f_0}}$ are estimated 
for the two values of  $g_{_{f_0 NN}},  g_{_{a_0 NN}} $ shown 
in Eq.~(\ref{sNN}).
We note that the contributions of  
$\eta_{hV}^{(0)}$ and  $\eta_{rS}^{(3)}$ are significantly enhanced in the MEM
%$\sim$2.4 
in comparison with the DNM. 
Also, the heavy $J/\Psi$ and $\Upsilon$ 
mesons involve  charmed and bottom quarks to contribute to \m-conversion 
via vector currents. This effect is absent in the DNM. 
Thus taking into account the MEM allows setting new limits from \m-conversion 
on the parameters of the underlying LFV models beyond the SM. 

\section{Limits on LFV parameters from $\mu^{-}-e^{-}$-conversion}
\label{limits}

%Using the standard procedure, we can derive from  
%Lagrangian~(\ref{eff-N}) the formula for the branching ratio of the coherent 
%$\mu^- - e^-$ conversion. 
%To leading order in the non-relativistic  reduction 
The branching ratio of $\mu^- - e^-$ conversion can be written in the form ~\cite{Kosmas:2000nj,Kosmas:1993ch}:
\begin{equation} 
R_{\mu e^-}^{coh} \ = \  
\frac{{\cal Q}} {2 \pi \Lambda_{LFV}^4} \  \   
\frac{p_e E_e \ ({\cal M}_p + {\cal M}_n)^2 } 
{ \Gamma_{\mu c} } 
\, , 
\label{Rme}
\end{equation} 
with $p_e, E_e$ being 3-momentum and energy of the final electron, 
respectively. Here 
${{\cal M}}_{p,n}$ are the nuclear $\mu^- - e^-$  transition matrix elements 
and $\Gamma_{\mu c}$ is the total rate of the ordinary muon capture. 
The factor ${\cal Q}$ can be expressed in terms of the parameters of Lagrangian~(\ref{eff-N}) as
\begin{eqnarray}
\hspace*{-1cm}
{\cal Q} &=& |\alpha_{VV}^{(0)}+\alpha_{VV}^{(3)}\ \phi|^2 +
|\alpha_{AV}^{(0)}+\alpha_{AV}^{(3)} \phi|^2 + 
|\alpha_{SS}^{(0)}+\alpha_{SS}^{(3)} \phi|^2 + 
|\alpha_{PS}^{(0)} + \alpha_{PS}^{(3)} \phi|^2 
\nonumber  \\ 
\hspace*{-1cm} 
&+& 2{\rm Re}\{(\alpha_{VV}^{(0)}+\alpha_{VV}^{(3)} 
\phi)(\alpha_{SS}^{(0)}+ \alpha_{SS}^{(3)} \phi)^\ast
+  (\alpha_{AV}^{(0)}+\alpha_{AV}^{(3)}\ \phi)(\alpha_{PS}^{(0)} + 
\alpha_{PS}^{(3)}\ \phi)^\ast\}\, .
\label{Rme.1} 
\end{eqnarray} 
This expression involves the nuclear structure factor
\begin{eqnarray}\label{phi}
\phi = ({\cal M}_p - {\cal M}_n)/({\cal M}_p + {\cal M}_n).
\end{eqnarray}
%which is typically rather small for the experimentally interesting nuclei.

The nuclear matrix elements ${\cal M}_{p,n}$ have been calculated 
in Refs.~\cite{Kosmas:2001mv,Faessler:2000pn} for 
$^{27}$Al, $^{48}$Ti, $^{197}$Au and $^{208}$Pb. 
Their values are presented in Table~\ref{Table-1} 
together with data for the total rates $\Gamma_{\mu c}$ of 
ordinary muon capture~\cite{Suzuki:1987jf} and the 3-momentum $p_e$ of 
the final electron.

\begin{table}

%%%%%%%%%%%%%%%%%%%%%%%%%%%%%%%%%%%%%%
\begin{center}
\begin{tabular}{|r|c|c|c|c|}
\hline
  &  &  &  &    \\
Nucleus & $p_e \, (fm^{-1})$ &
$\Gamma_{\mu c} \, ( \times 10^{6} \, s^{-1})$ &
${\cal M}_p(fm^{-3/2})$ & ${\cal M}_n(fm^{-3/2})$  \\
\hline
  &  &  &  &    \\
$^{27}$Al  & 0.531 &   0.71 & 0.047 & 0.045  \\
  &  &  &  &   \\
$^{48}$Ti  & 0.529 &  2.60 & 0.104 & 0.127   \\
   &  &  &  &  \\
$^{197}$Au & 0.485& 13.07 & 0.395 & 0.516   \\
  &  &  &  &    \\ 
$^{208}Pb$ & 0.482 & 13.45 & 0.414 & 0.566    \\
\hline
\end{tabular}
\end{center}
\caption{\m $ $ nuclear matrix elements, ${\cal M}_{p,n}$ ,
and other quantities from  Eqs. (\ref{Rme}).}
\label{Table-1}
\end{table}
%%%%%%%%%%%%%%%%%%%%%%%%%%%%%%%%%%%%%%

With the parameters  from Table~\ref{Table-1} we find that the presently 
most stringent limits on the dimensionless lepton-nucleon LFV couplings 
$\alpha$ of the Lagrangian (\ref{eff-N}) result from the SINDRUM II searches 
for \m-conversion on $^{198}$Au~\cite{Bertl:2006up}. 
Here we show these limits and the limits corresponding to the future
experiment PRISM/PRIME  \cite{Witte:2012zza} with titanium $^{48}$Ti  target
aiming at the sensitivity of $10^{-18}$. We have for these two cases:  
\begin{eqnarray}\label{alpha-lim-Au}
&&R_{\mu e}^{Au} \leq 4.3 \times 10^{-12} \  \  \mbox{\cite{Bertl:2006up}}:   \ \ \ \alpha_{h V, rS}^{(k)} \left(\frac{1 \mbox{GeV}}{\Lambda_{LFV}}\right)^2 \leq 
8.5\times 10^{-13} B^{(k)}(Au),\\
\label{alpha-lim-Ti}
&&R_{\mu e}^{Ti}\  \lsim 10^{-18} \ \ \  \ \ \ \ \  \mbox{\cite{Witte:2012zza}}:   \ \ \ \alpha_{h V, rS}^{(k)} \left(\frac{1 \mbox{GeV}}{\Lambda_{LFV}}\right)^2 \leq 
1.6 \times 10^{-15} B^{(k)}(Ti),
\end{eqnarray}
with $k=0,3$, $h=A,V$, $r=S,P$ and  $B^{0,3}(Ti) = (1, 10)$, $B^{0,3}(Au) = (1,  7.5)$.

The limits in Eqs.~(\ref{alpha-lim-Au}) 
and (\ref{alpha-lim-Ti}) can be used for  derivation of  individual bounds 
on the terms contributing to the 
coefficients $\alpha $. Following the common practice we assume the absence of substantial
cancellations  between different terms. 
Thus, from Eqs.~(\ref{DNM-alpha-V}), (\ref{DNM-alpha-S}), 
(\ref{MEM-alpha-V}) and (\ref{MEM-alpha-S}) we deduce upper limits on the dimensionless couplings of 
%constraints on 
the 4-fermion quark-lepton LFV contact terms of the 
Lagrangian~(\ref{eff-q}) for the two studied mechanisms of hadronization: 
for the direct nucleon mechanism (DNM) and for the meson-exchange mechanism (MEM). 
These limits are listed in Tables~\ref{Table-2}, \ref{Table-3}. 
%
%%%%%%%%%%%%%%%%%%%%%%%%%%%%%%%%%%%%%%%
%
\begin{table}

\vspace*{.4cm} 

\begin{center} 
\begin{tabular}{|c|c|c||c|c|c|} 
\hline
&&&&&    \\[-3mm]
$\eta$ & DNM $\times \left(\frac{\Lambda_{LFV}}{1 \mbox{GeV}} \right)^{2}$ & MEM $\times \left(\frac{\Lambda_{LFV}}{1 \mbox{GeV}} \right)^{2}$& $
\Lambda_{\mu e}$ in TeV & DNM & MEM\\[3mm]
\hline 
&&&&&    \\[-3mm]
$\eta^{(3)}_{hV}$ & $1.3 \times 10^{-11}$ & $1.6 \times 10^{-11}$ & $\Lambda_{\mu e}^{(3) hV} $ & $10^{3}$ &   900    \\[1mm]
\hline
&&&&&    \\[-3mm]
$\eta^{(0)}_{hV}$& $5.7 \times 10^{-13}$ & $2.4 \times 10^{-13}$ & $\Lambda_{\mu e}^{(0) hV} $ & $4.7 \times 10^{3}$ & $7.2 \times 10^{3}$    \\[1mm]
\hline
&&&&&    \\[-3mm]
$\eta^{(s)}_{hV}$& No limits & $2.8 \times 10^{-11}; 4.3 \times 10^{-13}$ &$\Lambda_{\mu e}^{(s) hV} $ &  No limits &  $770; 5.4\times 10^{3} $   \\[1mm]
\hline
&&&&&    \\[-3mm]
$\eta^{(c)}_{hV}$& No limits & $4.3 \times 10^{-9}$ & $\Lambda_{\mu e}^{(c) hV} $ & No limits & 54   \\[1mm]
\hline
&&&&&    \\[-3mm]
$\eta^{(b)}_{hV}$& No limits & $1.6 \times 10^{-6}$ & $\Lambda_{\mu e}^{(b) hV} $& No limits & 3   \\[1mm]
\hline
&&&&&    \\[-3mm]
$\eta^{(3)}_{r S}$& $1.2 \times 10^{-11} [1.6 \times 10^{-11}]$ & $4.0 \times 10^{-12} [6.9 \times 10^{-12}]$ & $\Lambda_{\mu e}^{(3) rS} $ & $10^{3} [900]$ &
$1.8 \times 10^{3} [1.4 \times 10^{3}]$    \\[1mm]
\hline
&&&&&    \\[-3mm]
$\eta^{(0)}_{r S}$& $2.7 \times 10^{-13} [1.8 \times 10^{-13}]$ & $3.7 \times 10^{-13} [6.4 \times 10^{-13}]$ & $\Lambda_{\mu e}^{(0) rS} $ & 
$6.8 \times 10^{3} [8.4 \times 10^{3}]$& $5.8 \times 10^{3} [4.4 \times 10^{3}]$   \\[1mm]
\hline
&&&&&    \\[-3mm]
$\eta^{(s)}_{r S}$& $1.3 \times 10^{-12} [3.4 \times 10^{-13}] $ & No limits & $\Lambda_{\mu e}^{(s) rS} $ & $3\times 10^{3} [6 \times 10^{3}]$ & 
No limits    \\[1mm]
\hline
&&&&&    \\[-3mm]
$\eta^{(c)}_{r S}$& $1.4 \times 10^{-11}$ & No limits & $\Lambda_{\mu e}^{(c) rS} $ & 950 &   No limits   \\[1mm]
\hline
&&&&&    \\[-3mm]
$\eta^{(b)}_{r S}$& $4.3 \times 10^{-11}$ & No limits & $\Lambda_{\mu e}^{(b) rS} $ & 540 &  No limits    \\[1mm]
\hline
&&&&&    \\[-3mm]
$\eta^{(t)}_{r S}$& $1.7 \times 10^{-9}$ & No limits & $\Lambda_{\mu e}^{(t) rS} $ & 90 &  No limits    \\[1mm]
\hline
\end{tabular}
\end{center} 
\caption{
%{SK_22.03.2013}
%Lower 
Upper limits on the LFV parameters  $\eta$ of the quark-lepton 
contact operators in Eq. (\ref{eff-q}), and lower limits 
%{SK_22.03.2012} END
on their individual mass scales, 
$\Lambda_{\mu e}$, defined in Eq. (\ref{Lambda-LFV}), inferred from 
the SINDRUM II data for  $^{198}$Au \cite{Bertl:2006up}. We show the limits  
both for the direct nucleon mechanism (DNM) and the meson exchange 
mechanism (MEM).  For $\eta^{(s)}_{h V}$ 
in the 3rd column and for $\Lambda_{\mu e}^{(s)hV}$ in the last 
column we show the two limits corresponding to the upper and lower  
bounds of the interval for  $g_{\phi NN}$ in 
Eq.~(\ref{VN-couplings}). The limits in the square brackets correspond 
to the options shown in Eqs. (\ref{FFN}) and  (\ref{sNN}). }
\label{Table-2}
\end{table} 
%%%%%%%%%%%%%%%%%%%%%%%%%%%%%%%%%%%%%%%
%
%%%%%%%%%%%%%%%%%%%%%%%%%%%%%%%%%%%%%%%
%
\begin{table}

\vspace*{.4cm} 

\begin{center} 
\begin{tabular}{|c|c|c||c|c|c|} 
\hline
&&&&&    \\[-3mm]
$\eta$ & DNM $\times \left(\frac{\Lambda_{LFV}}{1 \mbox{GeV}} \right)^{2}$ & MEM $\times \left(\frac{\Lambda_{LFV}}{1 \mbox{GeV}} \right)^{2}$& $
\Lambda_{\mu e}$ in TeV & DNM & MEM\\[3mm]
\hline 
&&&&&    \\[-3mm]
$\eta^{(3)}_{hV}$ & $2.1 \times 10^{-14}$ & $2.6 \times 10^{-14}$ & $\Lambda_{\mu e}^{(3) hV} $ & $2.5\times 10^{4}$ &   $2.3 \times 10^{4}$    \\[1mm]
\hline
&&&&&    \\[-3mm]
$\eta^{(0)}_{hV}$& $1.1 \times 10^{-15}$ & $4.8 \times 10^{-16}$ & $\Lambda_{\mu e}^{(0) hV} $ & $9.4 \times 10^{4}$ & $1.4 \times 10^{5}$    \\[1mm]
\hline
&&&&&    \\[-3mm]
$\eta^{(s)}_{hV}$& No limits & $5.6 \times 10^{-14}; 8.6 \times 10^{-16}$ &$\Lambda_{\mu e}^{(s) hV} $ &  No limits &  $1.5 \times 10^{4} ; 1.1 \times 10^{5} $   \\[1mm]
\hline
&&&&&    \\[-3mm]
$\eta^{(c)}_{hV}$& No limits & $8.6 \times 10^{-12}$ & $\Lambda_{\mu e}^{(c) hV} $ & No limits & $1.1 \times 10^{3}$   \\[1mm]
\hline
&&&&&    \\[-3mm]
$\eta^{(b)}_{hV}$& No limits & $3.2 \times 10^{-9}$ & $\Lambda_{\mu e}^{(b) hV} $& No limits & 60   \\[1mm]
\hline
&&&&&    \\[-3mm]
$\eta^{(3)}_{r S}$& $2.0 \times 10^{-14} [2.6 \times 10^{-14}]$ & $6.4 \times 10^{-15} [1.1 \times 10^{-14}]$ & $\Lambda_{\mu e}^{(3) rS} $ & $2.5 \times 10^{4} [2.3 \times 10^{4}]$ &
$4.5 \times 10^{4} [3.5 \times 10^{4}]$    \\[1mm]
\hline
&&&&&    \\[-3mm]
$\eta^{(0)}_{r S}$& $5.4 \times 10^{-16} [3.6 \times 10^{-16}]$ & $7.4 \times 10^{-16} [1.3 \times 10^{-15}]$ & $\Lambda_{\mu e}^{(0) rS} $ & 
$1.4 \times 10^{5} [1.7 \times 10^{5}]$& $1.2 \times 10^{5} [8.8 \times 10^{4}]$   \\[1mm]
\hline
&&&&&    \\[-3mm]
$\eta^{(s)}_{r S}$& $2.6 \times 10^{-15} [6.8 \times 10^{-15}] $ & No limits & $\Lambda_{\mu e}^{(s) rS} $ & $6\times 10^{4} [1.2 \times 10^{5}]$ & 
No limits    \\[1mm]
\hline
&&&&&    \\[-3mm]
$\eta^{(c)}_{r S}$& $2.8 \times 10^{-14}$ & No limits & $\Lambda_{\mu e}^{(c) rS} $ & $1.9 \times 10^{4}$ &   No limits   \\[1mm]
\hline
&&&&&    \\[-3mm]
$\eta^{(b)}_{r S}$& $8.6 \times 10^{-14}$ & No limits & $\Lambda_{\mu e}^{(b) rS} $ & $1.1 \times 10^{4}$ &  No limits    \\[1mm]
\hline
&&&&&    \\[-3mm]
$\eta^{(t)}_{r S}$& $3.4 \times 10^{-12}$ & No limits & $\Lambda_{\mu e}^{(t) rS} $ & $1.8 \times 10^{3}$ &  No limits    \\[1mm]
\hline
\end{tabular}
\end{center} 
\caption{The same as in Table~\ref{Table-2} but for the expected sensitivities of the future experiment 
PRISM/PRIME  \cite{Witte:2012zza} with titanium $^{48}$Ti.}
\label{Table-3}
\end{table} 
%%%%%%%%%%%%%%%%%%%%%%%%%%%%%%%%%%%%%%%

In Tables~\ref{Table-2}, \ref{Table-3} we also show lower limits 
on the individual mass scales, $\Lambda^{ij}_{\mu e}$, 
of the quark-lepton contact operators in Eq.~(\ref{eff-q}). 
In the conventional definition these scales are
related to our notations as 
\begin{eqnarray}\label{Lambda-LFV}
|\eta_{z}^{(a)}| \, \left(\frac{1{\rm GeV}}{\Lambda_{LFV}}\right)^2 = 
4\pi \left(\frac{1{\rm GeV}}{\Lambda^{(a) z}_{\mu e}}\right)^2
\end{eqnarray}
with $a=0, 3, s, c,b, t$ and $z= hV, rS$, where $h = A, V$ and $r = P, S$ 
as defined before.   

Let us compare our limits for the mass scales $\Lambda_{\mu e}$ with 
similar limits existing in the literature. 
It is a custom to refer to $\pi^-\rightarrow e^-\nu_e$ 
as the process which provides the most stringent limits on the lepton 
flavor conserving contact terms involving  pseudoscalar and scalar quark currents~\cite{Shanker:1982nd}. 
Note, the latter does not contribute directly to this process, 
but due to the gauge invariance with respect to the SM group one 
can relate the couplings of the scalar and pseudoscalar 
lepton-quark contact operators. 
%Hence the scalar contact terms are related to $\pi^-\rightarrow e^-\nu_e$.   
The updated upper limit from 
$\pi^-\rightarrow e^-\nu_e$~\cite{Particle-Data-Group:2012uq}  on the corresponding  mass scale is
\mbox{$\Lambda_{e e} \geq 500 \mbox{TeV}$.}
This limit is not related to our limits for the LFV mass scales 
$\Lambda_{\mu e}$. However, it can be taken as a reference value, 
illustrating the present situation with the (pseudo-)scalar contact terms.
Limits  on $\Lambda_{\mu e}$  of the LFV (pseudo-)scalar contact terms were  derived  in the literature from the experimental 
bounds on $\pi^+\rightarrow \mu^+ \nu_e$, 
$\pi^0\rightarrow \mu^\pm e^\mp$~\cite{Kim:1997rr}. 
Typical limits from these processes are  
$\Lambda_{\mu e} \geq  \mbox{few TeV}$.

As to the vector lepton-quark contact interactions, the corresponding scales 
can be extracted from the experimental 
limits~\cite{Particle-Data-Group:2012uq} on 
$M_{V} \rightarrow \mu^{\pm} e^{\mp}$ for $M_{V} 
= \rho, \omega, \phi, J/\Psi, \Upsilon$. 
But, in Refs.~\cite{Gutsche:2009vp,Gutsche:2011bi} it was shown 
that \m-conversion is much more sensitive probe of  the LFV physics 
%sets much better limits on these contact interactions 
than vector meson decays. Therefore limits on $\Lambda_{\mu e}$ 
from \m-conversion must be much better than limits from these decays.

Recently the ATLAS  Collaboration reported results of  an analysis of 
Drell-Yan $e^{-}e^{+}$ and $\mu^{-} \mu^{+}$ dileptons from the data 
collected in 2011 at the LHC  with $\sqrt{s} = 7$ TeV \cite{Aad:2011tq}.  
They set lower limits on the scale of the lepton flavor conserving 
lepton-quark vector contact interactions with the typical values 
$\Lambda_{ee} \gsim 10$ TeV and $\Lambda_{\mu\mu} \gsim 5$ TeV .   
The LHC experiments are also able to constrain the LFV lepton-quark contact 
interaction scales $\Lambda_{l l'}$ from the measurement of Drell-Yan cross 
sections in the high dilepton mass region~\cite{Krasnikov:2003ef}. 
In this case typical expected limits are $\Lambda_{l l'} \geq 35 \mbox{TeV}$. 
A comparison of the above mentioned  limits existing in the literature with the ones in Table~\ref{Table-2}, 
extracted from \mbox{\m conversion}, shows that our limits  are more stringent 
with the only possible exception of the scale of the $bb\mu e$  vector contact 
interaction.  
However, as seen from Table~\ref{Table-3}, the future PRISM/PRIME 
experiment~\cite{Witte:2012zza} would be able to set such limits on the 
scales of  all the contact LFV interactions of the type $qq\mu e$ that look hardly accessible for other experiments including those at the LHC.

\section{Summary} 

In this paper we analyzed the nuclear $\mu^--e^-$ conversion using 
general framework  of effective Lagrangians without 
referring to any particular LFV model beyond the SM.
We examined two hadronization mechanisms of the underlying effective 
quark-lepton LFV Lagrangian (\ref{eff-q}): the direct nucleon (DNM) and 
the meson exchange (MEM) mechanisms. 
%
%We showed that a role of the MEM mechanism is comparable 
%with the DNM mechanism. In some cases MEM contribution is even larger.  
%Thus, taking meson-exchange into account modifies the limits on the LFV 
%lepton-quark couplings derived on the basis of the conventional direct 
%nucleon mechanism making  

Using experimental upper bounds on the \m conversion rate we extracted 
lower limits on the mass scales $\Lambda_{\mu e}$ of the LFV lepton-quark contact vector 
and scalar terms  $qq \mu e$ involved in this process for all quark 
flavors $q=u,d,s,c,b,t$. 
We showed that these limits are more stringent than the similar ones 
existing in the literature, including the limits from the present 
experimental data on meson decays and the limits expected from 
the future experiments at the LHC.
 
We demonstrated that neither of the two hadronization mechanisms, DNM and MEM, should be overlooked in analysis of \m conversion
due to their complementarity.  As seen from Tables II and III in some cases it is the DNM which is only able to set limits on 
%only with the DNM is possible to set limits on 
the corresponding LFV parameters while in some other cases it is the MEM.
%this can be done only on the basis of the MEM.
Also MEM improves the limits on the LFV parameters $\Lambda^{(0) hV}_{\mu e}$ and $\Lambda^{(3) rS}_{\mu e}$ in comparison with
the conventional DNM mechanism. This fact may have an appreciable impact on the phenomenology of 
the LFV physics beyond the Standard Model.

\begin{acknowledgments}

This work was supported by the FONDECYT projects 1100582, 1100287, 11121557,   by \mbox{CONICYT} within the  
Centro-Cient\'\i fico-Tecnol\'{o}gico de Valpara\'\i so PBCT ACT-028,
 by the DGIP of the UTFSM and 
by the  DFG under Contract No. LY 114/2-1. 
%The work was also done partially under 
A partial support was also recieved from 
the project 2.3684.2011 of 
Tomsk State University.  
V. E. L. would like to thank Departamento de F\'\i sica y Centro
Cient\'\i fico Tecnol\'ogico de Valpara\'\i so (CCTVal), Universidad
T\'ecnica Federico Santa Mar\'\i a, Valpara\'\i so, Chile for warm
hospitality. 

\end{acknowledgments}

\end{document}